\documentclass[copyright]{eptcs}

\providecommand{\event}{MARS 2017}

\usepackage{underscore}
\usepackage{breakurl}
\usepackage{graphicx}

\usepackage{xcolor}
\definecolor{grey}{cmyk}{0,0,0,0.8}
\hypersetup{bookmarks=true,bookmarksopen=false,colorlinks=true,linkcolor=grey,citecolor=grey,urlcolor=grey}

\usepackage{listings}

\usepackage{cadp-lnt}

\newcommand{\T}[1]{\mbox{\lstinline+#1+}}

\title{The Unheralded Value of the Multiway Rendezvous: Illustration with the Production Cell Benchmark}

\def\titlerunning{The Unheralded Value of the Multiway Rendezvous}

\author{Hubert Garavel \quad\qquad Wendelin Serwe
  \institute{INRIA Grenoble, France}
  \institute{Univ. Grenoble Alpes, LIG, F-38000 Grenoble, France}
  \institute{CNRS, LIG, F-38000 Grenoble, France}
  \email{Hubert.Garavel@inria.fr \quad\qquad Wendelin.Serwe@inria.fr}
}

\def\authorrunning{H.~Garavel \& W.~Serwe}

\begin{document}
\maketitle

\sloppy

\begin{abstract}
The multiway rendezvous introduced in Theoretical CSP is a powerful
paradigm to achieve synchronization and communication among a group of
(possibly more than two) processes. We illustrate the advantages of this
paradigm on the production cell benchmark, a model of a real metal
processing plant, for which we propose a compositional software 
controller, which is written in LNT and LOTOS, and makes intensive
use of the multiway rendezvous.
\end{abstract}

\section{Introduction}
\label{sec:introduction}

	We investigate the design of software controllers for complex systems. Concurrency is a natural way to specify such controllers by decomposing their software into separate processes, each dedicated to a specific activity or a specific aspect of the system. For instance, an automatic pilot may include two concurrent processes that control roll and pitch, respectively; also, a controller for a robot operating in a space with $n$ degrees of freedom may contain $n$ processes, each supervising the robot motion within a given degree of freedom.

	Concurrency is a high-level specification paradigm that can be implemented in diverse ways. On the one hand, implementations can be done in hardware, in software, or in a combination of both. On the other hand, implementations may either preserve the concurrency present at the specification level by translating it into parallel code, or remove concurrency by expanding/flattening it into sequential code.

	Whatever implementation techniques are chosen, the different processes that constitute a controller, even if they can be independent to a large degree, must also synchronize, communicate, and co-operate to achieve common goals and enforce global constraints applying to the system. Among the various paradigms proposed for synchronization and communication, the multiway rendezvous designed for Theoretical~CSP \cite{Brookes-Hoare-Roscoe-84} \cite{Hoare-85} \cite{Roscoe-Hoare-Bird-97} presents major advantages, although these are not always perceived or put forward.

	In this article, we illustrate the merits of the multiway rendezvous on a benchmark that once enjoyed a large visibility among the formal methods community: the production cell case study \cite{Lewerentz-Lindner-95-a}. For this benchmark, we developed a software controller, which makes intensive use of the multiway rendezvous and enjoys a nicely distributed architecture. This controller was first designed in LOTOS \cite{ISO-8807}, then in LNT \cite{Champelovier-Clerc-Garavel-et-al-10-v6.4}. The full code of the LNT specification, which is easier to read, is given in Appendix~\ref{sec:lnt}, but most of the discussion applies to both LOTOS and LNT.

	The remainder of this article is organized as follows.
	Section~\ref{sec:multiway} recalls the principles and benefits of the multiway rendezvous.
	Section~\ref{sec:production-cell} describes the production cell case study, an overview of formal specifications already developed for this benchmark being given in Appendix~\ref{sec:related-work}.
	Section~\ref{sec:model} presents the principles and the architecture of our LOTOS and LNT specifications.
	Section~\ref{sec:code-generation} details how controller implementations can be generated automatically from these specifications
	and Section~\ref{sec:validation} discussses validation issues.
	Finally, Section~\ref{sec:conclusion} gives a few concluding remarks.

\section{The Multiway Rendezvous}
\label{sec:multiway}

	From an historical point of view, the multiway rendezvous is not a concept designed in one day, but rather the result of a long evolution alternating major shifts and incremental improvements:

\begin{itemize}
	\item From the origins to the mid-70s, interprocess communication was mostly achieved using shared variables, whereas synchronization between concurrent processes relied upon memory-based mechanisms (semaphores, locks, critical sections, etc.). Such approaches had several drawbacks: lack of abstraction, existence of multiple incompatible semantics, difficulty to design correct programs, and difficulty for automated tools to analyze processes in which variables can be modified by other processes at any time.

	\item In 1978, C.A.R.~Hoare introduced CSP \cite{Hoare-78}, a language built around the concept of {\em rendezvous}, a new {\em message-passing\/} paradigm unifying synchronization and communication. A key advantage of this paradigm is that the few places where a variable can be modified by a concurrent process are explicitly documented. In CSP, a rendezvous can only take place between two processes (a sender and a receiver) and the parallel architecture is ``hard-coded'', as each process must explicitly indicate, at each rendezvous point, the name of the concurrent process it communicates with. 

	\item In 1980, R.~Milner proposed CCS \cite{Milner-80}, a language that reuses the concept of binary rendezvous, for which he defined a formal semantics. CCS solves the aforementioned issue with CSP by introducing the concept of {\em port\/} that allows for reusable process components and parameterized parallel architectures. In CCS, processes no longer refer directly to other processes but only indirectly, using ports, which are intermediate communication objects that connect processes together.

	\item In 1984, S.D.~Brookes, C.A.R.~Hoare, and A.W.~Roscoe designed a refined version of CSP named TCSP ({\em Theoretical CSP\/}) \cite{Brookes-Hoare-Roscoe-84} \cite{Hoare-85}, which combines ideas from CSP and CCS. A major innovation brought by TCSP is the {\em multiway rendezvous}, which generalizes binary rendezvous to more than two processes. A formal semantics (given in terms of traces and refusals) takes care of the presence of multiple senders and/or receivers.

	\item At the same time, an ISO standardization committee headed by E.~Brinksma had undertaken the definition of LOTOS, a new formal language to describe communication protocols. The committee initially selected the binary rendezvous of CCS, until A.J.~Tocher presented the TCSP multiway rendezvous, which was adopted and included in the standard \cite{ISO-8807}. LOTOS brought useful features, such as multiple value parameters, strict type checking, and the extension of {\em selection predicates} (i.e., Boolean guards that forbid rendezvous if they evaluate to false) to multiway rendezvous.

	In our opinion, multiway rendezvous is one of the best features of LOTOS, while none of the two other standards, Estelle \cite{ISO-9074} and SDL \cite{CCITT-Z100} that competed with LOTOS at those times, provided a similar expressiveness. It is therefore no surprise that multiway rendezvous has been preserved in the next-generation languages based on LOTOS, namely E-LOTOS \cite{ISO-15437} and LNT \cite{Champelovier-Clerc-Garavel-et-al-10-v6.4}, as well as in the FDR2 \cite{FDR2-10} implementation of TCSP.

	\item In 1999, H.~Garavel and M.~Sighireanu proposed ``graphical'' parallel composition operators \cite{Garavel-Sighireanu-99}, which take $n$ arguments (whereas the traditional parallel composition operators accept only two arguments) and are thus better in line with the concept of multiway rendezvous. These operators have been implemented in LNT and are used in Appendix~\ref{sec:lnt} of the present article.
\end{itemize}

	Beyond these languages, the multiway rendezvous paradigm has not spread as largely as one would wish. One reason for this is the influence of CCS, which promotes an incompatible paradigm of binary-only communication. Another reason is the high difficulty to properly implement multiway rendezvous, either in a sequential setting or in a distributed setting; for the latter point, which is even more difficult, let us mention recent work that implements the LOTOS multiway rendezvous among a collection of distributed processes interconnected by POSIX sockets \cite{Evrard-Lang-15} \cite{Evrard-Lang-13} \cite{Evrard-16}.
However, concepts similar or close to the multiway rendezvous are indeed present in certain computer languages or models:

\begin{itemize}
	\item The mCRL2 process algebra \cite{Groote-Mousavi-14} also contains multiway synchronization. Compared to LOTOS and CSP, the main difference is that mCRL2 actions are ``output-only'' (at least syntactically), because mCRL2 does not feature the CSP notations for inputs and outputs: ``\T{?}'' is absent and ``\T{!}'' is implicit --- see \cite[Section~3.3]{Garavel-15-b} for details.

	\item Petri nets can naturally express multiway synchronization between $n \geq 2$ processes by means of transitions having $n$ input places and $n$ output places. The C{\AE}SAR compiler~\cite{Garavel-Sifakis-90}, which translates LOTOS terms to interpreted Petri nets, uses this Petri-net feature to implement LOTOS multiway rendezvous.

	\item Barriers are lower-level mechanisms to collectively synchronize a set of  processes or threads. The multiway rendezvous can be seen as a powerful generalization of barriers with: (i) data exchange capabilities taking place when all the processes/threads have reached the barrier, and (ii) the possibility for a process to choose between different barriers.

	\item Synchronous languages also possess related concepts. For instance, Esterel \cite{Berry-Gonthier-92} \cite{PotopButucaru-Edwards-Berry-07} can synchronize $n \geq 2$ actions and compose together the values carried by each of these actions. To a certain extent, the multiway rendezvous imports synchronous concepts into an asynchronous setting: one can indeed use the multiway rendezvous to force a set of concurrent processes to synchronize, and possibly exchange values at every tick of some logical clock.
\end{itemize}

	In spite of the implementation difficulties, the multiway rendezvous remains a natural way to express synchronization among a set of distributed processes, as well as an irreplaceable mechanism to describe certain situations that, even if less frequent than binary communication, are not uncommon. Four examples of such situations are:

\begin{itemize}
	\item {\em Observers\/}: It is often useful to monitor data exchanges between two communicating processes. For instance, one may wish to count the number of messages exchanged between these processes or to build the list of such messages. This is not easy in languages that rely on binary communication, and even impossible in the case of CCS, where the synchronization of an emission and a reception is immediately turned into a $\tau$ (i.e., invisible or almost invisible) action. On the contrary, multiway rendezvous makes it easy to introduce a third ``observer'' process that also synchronizes on the communication action using a three-party rendezvous, without perturbing the two other processes.

	\item {\em Supervisors\/}: A step beyond observers is to introduce a third ``supervisor'' process that not only observes communications passively, but also actively interferes by allowing or blocking certain communications, depending on the communication contents and/or the internal state of the supervisor process. For instance, a supervisor process may serialize actions by forcing them to occur in a specified order. This is easy to achieve using multiway synchronization, as a rendezvous can only take place when all participants (including the supervisor) agree.

	An extended form of supervision is the {\em constraint-oriented\/} specification style \cite{Vissers-Scollo-vanSinderen-88} \cite{Vissers-Scollo-vanSinderen-Brinksma-91}, in which each process imposes its specific constraints over exchanged data values or action order. Putting all these processes in parallel using multiway rendezvous amounts to taking the logical conjunction of all the constraints expressed by these processes. The execution of such a parallel composition behaves like a constraint solver that searches for possible solutions, if any.

	\item {\em Consensus\/}: The multiway rendezvous between $n$ processes is a powerful abstraction that achieves, in a single atomic operation, a distributed consensus protocol. Describing the same protocol using binary communications is likely to cause an exponential blow-up, as all possible interleavings between the actions of the $n$ processes may occur. A salient example can be found in \cite[Section~3]{Garavel-Hermanns-02}, where multiway rendezvous is used to model the arbitration mechanism of the SCSI-2 hardware bus. In this example, an eight-party rendezvous expresses, in one atomic action: (i) a voting procedure in which each hardware device declares whether it wants to access the bus or not; (ii) the selection, among all devices requesting access, of the device with the lowest number; and (iii) the notification to each device whether it was granted access or not.

	\item {\em Coordination\/}: In the present article, we illustrate yet another application of the multiway rendezvous. Given a software controller for a system evolving in a space with $n$ degrees of freedom, each degree being managed by a separate concurrent process in the controller, we use the multiway rendezvous to express high-level coordination goals between these processes, such as moving from a starting point $A$ to a target point $B$. Each process is responsible for moving along one axis; depending on their respective speed, the various processes may reach their target in a nondeterministic order. Therefore, point $B$ is only reached when all processes have individually reached their target, which is conveniently expressed by a multiway rendezvous between synchronizing the $n$ processes. As a side remark, we only use the synchronization capabilities of the multiway rendezvous, as the problem requires no exchange of values when multiway rendezvous take place.

\end{itemize}

\section{The Production Cell Case Study} 
\label{sec:production-cell}

The case study ``Control Software for an Industrial Production Cell''~\cite{Lewerentz-Lindner-95-a-introduction} \cite{Lewerentz-Lindner-95-b} was proposed in the 90s as a benchmark to assess the benefits of different formal methods applied to a common critical software system.
The task description~\cite{Lindner-95} required to use a formal method to develop a software controller for a production cell, replicating a real metal processing plant in Karlsruhe, Germany.
The benchmark became popular and, in 1995, a book devoted to the production cell case study was published~\cite{Lewerentz-Lindner-95-a}.

\subsection{Overview of the Production Cell}

The production cell operates on metal plates (or \emph{blanks}), which are brought into the cell by a \emph{feed belt}, transported to a \emph{press} via an \emph{elevating rotary table} and a two-armed \emph{robot}, before they leave the cell on the \emph{deposit belt}.
Diverging from the concrete production cell and to obtain a cyclic behaviour, the blanks are transported by a \emph{crane} from the deposit back to the feed belt.

The production cell is controlled by thirteen \emph{actuators} $A_1$, ..., $A_{13}$ (motors and magnets) and is equipped with fourteen \emph{sensors} $S_1$, ..., $S_{14}$ (switches, potentiometers, and photoelectric cells) to deliver status information to the controller.

\subsection{The Graphical Simulator of the Production Cell}

\begin{figure}
  \centering
  \includegraphics[width=.9\textwidth]{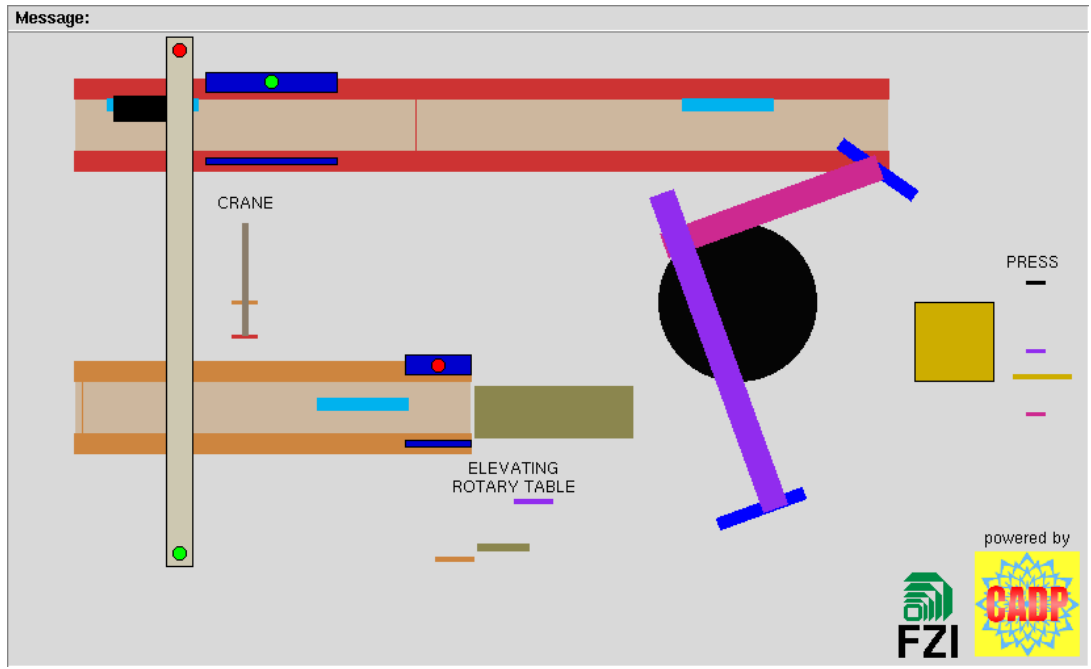}
  \caption{Screenshot of the graphical simulator}
  \label{fig:simulator-screenshot}
\end{figure}

	A graphical simulator~\cite{Brauer-Lewerentz-Lindner-93} \cite{Brauer-Lindner-95}, written in Tcl/Tk, enables prototype controllers to be validated and provides a reference to compare the controllers obtained from executable formal methods.
	Unfortunately, the Tcl/Tk source code of this simulator is no longer available today, as the FTP server \texttt{ftp.fzi.de} hosting the original version of the simulator does not seem to respond any more.
	Luckily, a copy of the simulator was archived at INRIA Grenoble, improved in a few points, and regularly adapted to the latest versions of Tcl/Tk and operating systems (Linux, MacOS, Windows, etc.).
	A screenshot of this simulator is shown in Figure~\ref{fig:simulator-screenshot}.

	The simulator has two functioning modes: {\em asynchronous\/} or {\em synchronous}. In principle, the asynchronous mode, which is {\em event-driven}, should be more efficient, as the production cell controller does not poll periodically the sensors and, thus, avoids busy-waiting loops (i.e., reading the value of sensors when this is not necessary). Alas, the \T{new_guard} command, which allows actions to be triggered when certain conditions (depending on sensor values) become true, does not seem to function, and its usage is discouraged by the authors of the simulator~\cite[Section~A.5.9]{Brauer-Lindner-95}.

	Therefore, the synchronous mode, which is is {\em cycle-driven}, remains the only available option. This mode is activated via the command-line option \T{-snc} and achieves bidirectional communication between the simulator and its controller via a simple protocol based on character-string commands and replies.
	In this mode, the production cell controller is expected to perform an infinite loop of successive {\em reaction steps}, i.e., periodically:
(i) acquire the current values of all sensors by sending a \T{get_status} command to the simulator;
(ii) compute the appropriate reaction;
(iii) send a sequence of commands to the actuators (at most one command per actuator); and
(iv) terminate the current reaction step by sending a \T{react} command that instructs the simulator to update its state by executing all received actuator commands.

\subsection{Prior Work on the Production Cell}
\label{sec:prior-work}

	The literature about the production cell case study is abundant. The reference book~\cite{Lewerentz-Lindner-95-a} describes the application of 18~different formal methods to the production cell case study. It then provides a brief comparative survey of these experiments~\cite{Lewerentz-Lindner-95-a-comparison} \cite{Lewerentz-Lindner-95-b}. 
	Since then, further experiments with other formal methods have been published separately. Some approaches also extend the original task description, e.g., \cite{Xu-Randell-Romanovsky-et-al-02} which investigates fault-tolerance, and \cite{Gondal-12}, which considers a production cell with two presses.

	In total, nearly 28~different formal methods have been applied to the production cell case study. Appendix~\ref{sec:related-work} of the present article gives two overview tables providing bibliographic references. Unfortunately, most of the source specifications are no longer available, so that it is difficult to discuss their characteristics in detail and make a precise comparison between them.

	Because the production cell benchmark comes with a graphical simulator, it clearly calls for {\em executable\/} formal methods, i.e., those from which executable code can be generated automatically and connected to the simulator. However, only  five prior experiments with executable formal methods can be found in the literature, with three out of five experiments being done with synchronous languages (see Table~\ref{tab:connected-models} in Appendix~\ref{sec:related-work}).

	The present article rather explores the asynchronous side of executable formal methods. Although several approaches have specified the production cell controller as a set of distributed processes combined using rendezvous~\cite{Rischel-Sun-97} or alternative synchronization primitives, such as interface functions~\cite{Boerger-Mearelli-97} or coordinated atomic actions~\cite{Zorzo-Romanovsky-Xu-et-al-99}, there was no automatic code generation from these models. And, as far as we are aware, no prior approach uses multiway rendezvous.

\section{The LOTOS and LNT Specifications of the Production Cell Controller}
\label{sec:model}

	An early LOTOS specification of a controller for the production cell was developed in July~1994 by the first author, and revised in August 1994 to produce a second version taking advantage of the multiway rendezvous and enabled the automatic generation of controller implementation in C.
	Although there was a kind offer to submit this specification for publication in the reference book~\cite{Lewerentz-Lindner-95-a}, there were still technical problems in connecting the controller to the graphical simulator, so that the LOTOS chapter for the book was left unfinished, the LOTOS specification being only mentioned in the comparative survey that forms Chapter~3 of the book~\cite{Lewerentz-Lindner-95-a-comparison}.
	The matter was put aside until 1997, where a fully functional version with an operational connection to the simulator was achieved with the help of Mark Jorgensen and integrated as a demonstration example\footnote{\url{ftp://ftp.inrialpes.fr/pub/vasy/demos/demo_19}} to the CADP toolbox.
	In 2013, the specification was translated to LNT, mostly by the second author, who also simplified the LOTOS specification and improved its runtime performance. Both specifications have been further enhanced in 2017 when preparing the present article. The latest version of the LNT specification is provided in Appendix~\ref{sec:lnt}.

\subsection{Architectural Decomposition of the Controller}

	We first present the principles underlying the production cell controller in LOTOS and LNT.
	Rather than having a monolithic controller (which, because of its complexity, could not easily evolve if the production cell was modified or reorganized), it is desirable to design the controller in a modular way, by assembling simpler components together.

\begin{figure}[t]
  \centering
  \includegraphics{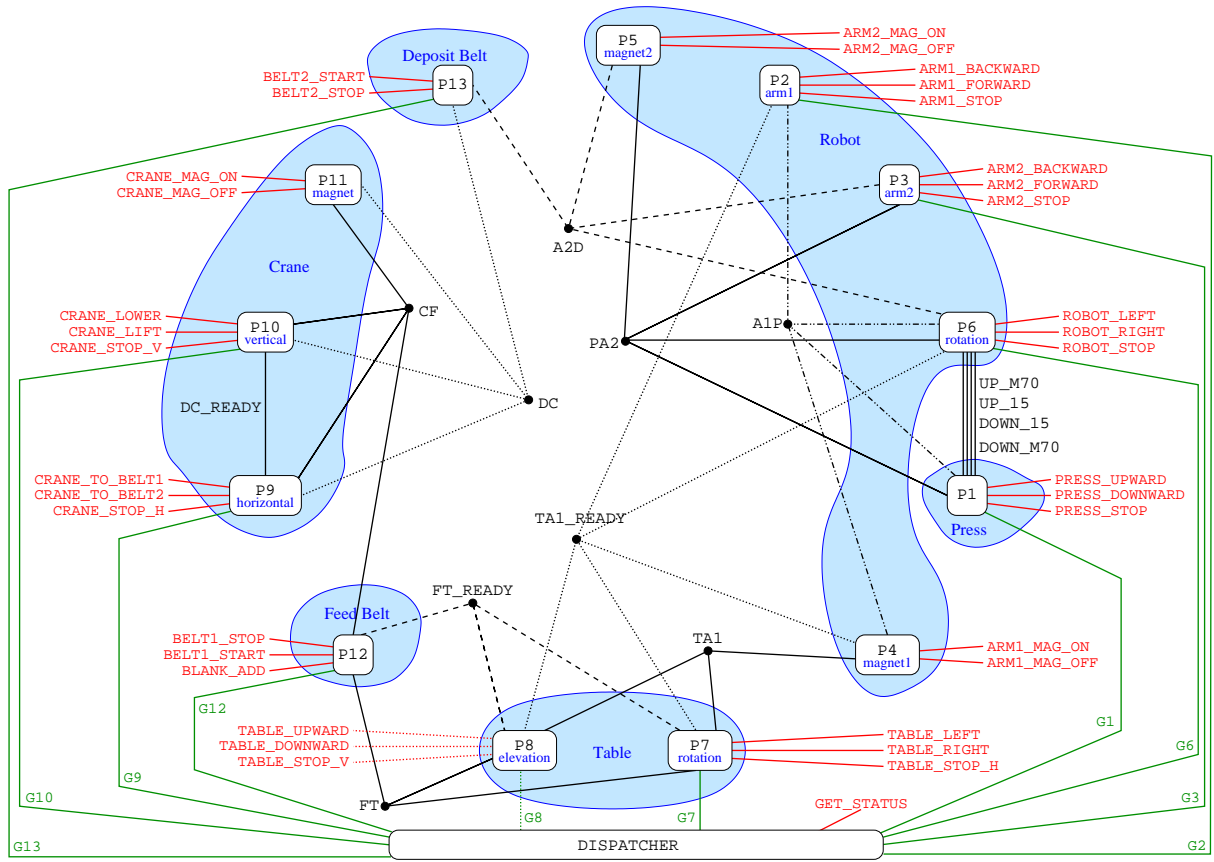}
  \caption{Architecture of the controller}
  \label{fig:controller-architecture}
\end{figure}

	The most natural way to decompose the controller is to follow the topology of the production cell, whose different devices (feed belt, rotary table, robot, deposit belt, crane) form a logical ring in which each device has to watch for its neighbours, with the additional fact that the robot and the press must communicate with each other.
	The overall architecture of the controller is illustrated in Figure~\ref{fig:controller-architecture}.

	More precisely, we choose to manage each separately controllable device (or degree of freedom of a device) of the production cell by a dedicated process. The controller can thus be decomposed into 13 concurrent processes \T{P1}, ..., \T{P13}, each process $\T{P}_i$ being in charge of the corresponding actuator $A_i$ (see Table~\ref{tab:processes} --- the indices of actuators are those given in \cite{Lindner-95} and \cite{Brauer-Lindner-95}).

\begin{table}
\begin{center}
\small
\begin{tabular}{cl} \hline
process & role \\ \hline
\T{P1} & move the lower part of the press vertically \\
\T{P2} & extend or retract the first robot arm \\
\T{P3} & extend or retract the second robot arm \\
\T{P4} & pick up or drop a metal plate with the first robot arm \\
\T{P5} & pick up or drop a metal plate with the second robot arm \\
\T{P6} & rotate the robot \\
\T{P7} & rotate the rotary table \\
\T{P8} & move the rotary table vertically \\
\T{P9} & move gripper of the travelling crane horizontally \\
\T{P10} & move gripper of the travelling crane vertically \\
\T{P11} & pick up or drop a metal plate with crane's gripper \\
\T{P12} & start or stop the motor of the feed belt \\
\T{P13} & start or stop the motor of the deposit belt \\ \hline
\end{tabular}
\end{center}
\caption{Processes of the production cell controller}
\label{tab:processes}
\end{table}

	Parallel composition is the natural way to express that the processes $\T{P}_i$ are largely independent from each other. Our controller is thus designed as a set of LOTOS and LNT processes that execute simultaneously and synchronize by rendezvous to coordinate those movements involving several devices.

	To each of the 34~protocol commands sent to actuators (e.g., \T{press_upward}, \T{press_stop}, etc.), we associate a corresponding LOTOS or LNT gate (named \T{PRESS_UPWARD}, \T{PRESS_STOP}, etc.) and we divide these gates into 13 groups numbered from 1 to 13, such that group $i$ contains the gates related to actuator $A_i$ \cite[Section~2.2.1]{Lindner-95}. The gate \T{BLANK_ADD} (corresponding to the command \T{blank_add}) is added to group~12 (feed belt) because new metal blanks are inserted into the production cell via the feed belt. Each process $\T{P}_i$ is responsible for accessing the gates of group $i$ and no other process $\T{P}_{j \neq i}$ can access these gates. Table~\ref{tab:gate-groups} lists the gates in each group.

\begin{table}
\begin{center}
\small
\begin{tabular}{cll} \hline
      group & \multicolumn{1}{c}{gates} & \multicolumn{1}{c}{actuator} \\
      \hline
      1 & \T{PRESS_UPWARD}, \T{PRESS_STOP}, \T{PRESS_DOWNWARD} & press \\
      2 & \T{ARM1_FORWARD}, \T{ARM1_STOP}, \T{ARM1_BACKWARD} & extension of arm~1 \\
      3 & \T{ARM2_FORWARD}, \T{ARM2_STOP}, \T{ARM2_BACKWARD} & extension of arm~2 \\
      4 & \T{ARM1_MAG_ON}, \T{ARM1_MAG_OFF} & magnet of arm~1 \\
      5 & \T{ARM2_MAG_ON}, \T{ARM2_MAG_OFF} & magnet of arm~2 \\
      6 & \T{ROBOT_LEFT}, \T{ROBOT_STOP}, \T{ROBOT_RIGHT} & robot rotation \\
      7 & \T{TABLE_LEFT}, \T{TABLE_STOP_H}, \T{TABLE_RIGHT} & table rotation \\
      8 & \T{TABLE_UPWARD}, \T{TABLE_STOP_V}, \T{TABLE_DOWNWARD} & table elevation \\
      9 & \T{CRANE_TO_BELT2}, \T{CRANE_STOP_H}, \T{CRANE_TO_BELT1} & move crane horizontally \\
      10 & \T{CRANE_LIFT}, \T{CRANE_STOP_V}, \T{CRANE_LOWER} & move crane vertically \\
      11 & \T{CRANE_MAG_ON}, \T{CRANE_MAG_OFF} & crane's magnet \\
      12 & \T{BELT1_START}, \T{BELT1_STOP}, \T{BLANK_ADD} & feed belt \\
      13 & \T{BELT2_START}, \T{BELT2_STOP} & deposit belt \\
      \hline
\end{tabular}
\end{center}
\caption{Gates grouped according to the controlled actuator}
\label{tab:gate-groups}
\end{table}

	Notice that the actuators commands could have been modelled differently by defining, rather than 35~gates without offer, only 13 gates (one per actuator) with output offers, i.e., values of enumerated types specifying the kind of movement expected (e.g., \T{UPWARD}, \T{STOP}, \T{DOWNWARD}, etc.). This solution was discarded because the previous one was simpler.

	The processes $\T{P}_i$ also have to synchronize together, and their interactions are dictated by the topology of the production cell. To achieve such synchronizations, the LOTOS and LNT specifications introduce 14~dedicated gates (named \T{FT_READY}, \T{FT}, \T{TA1_READY}, \T{TA1}, \T{A1P}, \T{PA2}, etc.) that remain internal to the controller.
	For instance, gate \T{PA2} expresses the (instantaneous) transfer of a blank from the press to the second arm of the robot. Other interactions are not instantaneous, so that processes need to synchronize at the beginning and at the end of the transfer; in such cases, two different gates are used. For instance, to transfer a blank from the feed belt to the table, the motor of the feed belt must be stopped when the blank arrives at its end (to avoid dropping the blank) until the table is correctly positioned (rendezvous on gate \T{FT_READY}) to receive the blank; then, the table must not move until the feed belt has been restarted (rendezvous on gate \T{FT}).

	In a few cases, rendezvous on these gates involve two processes only (e.g., \T{P9} and \T{P10} synchronize on gate \T{DC_READY}) but, usually, multiway rendezvous between three, four, or five processes is needed. 
	For instance, a three-party rendezvous on gate \T{FT} takes place when there is a blank at the end of the feed belt and the motor of the feed belt has been started (\T{P12}), and when the table is in a position (\T{P7} and \T{P8}) where it can receive a blank from the feed belt.
	A four-party rendezvous on gate \T{PA2} takes place when a blank is transferred from the press to the second arm of the robot and involves the process controlling the press (\T{P1}) and three processes controlling the robot for rotation (\T{P6}), extension of second arm (\T{P3}), and magnet of second arm (\T{P5}).
	A five-party rendezvous on gate \T{TA1_READY} takes place when the table is ready to deliver a blank element to the first arm of the robot and involves the two processes controlling the table (\T{P7} and \T{P8}), the two processes controlling the first arm of the robot (\T{P2} and \T{P4}), and the process controlling the rotation of the robot (\T{P6}).

	To interface this fully asynchronous controller with the simulator in synchronous mode, an additional \T{DISPATCHER} process was added, which acquires sensor values using an externally visible gate \T{GATE_STATUS} (corresponding to the \T{get_status} protocol command \cite{Brauer-Lindner-95}). Contrary to the commands for actuators, which can be emitted independently in any order, the sensor values must be acquired altogether (this is required by the protocol); thus, the dispatcher process is in charge of acquiring these values, as there is no logical criterion to select a particular process $\T{P}_i$ for this task.

	Then, the dispatcher process sends sensor values to each process $\T{P}_i$ (excepted the processes \T{P4}, \T{P5}, and \T{P11} that control the magnets) using a dedicated gate $\T{G}_i$. 
	There are no gates \T{G4}, \T{G5}, or \T{G11}, because the magnets have no related sensors and the moments at which magnets should be switched on or off can be determined by multiway rendezvous. For instance, the magnet of the second arm of the robot should be switched on when an item is delivered from the press to arm~2 (rendezvous on gate \T{PA2}) and switched off when an item is delivered from arm~2 to the deposit belt (rendezvous on gate \T{A2D}).

\subsection{Sensor Values and Data Abstractions}
\label{sec:model:data}

	To remain as close as possible to the notations given in \cite{Lindner-95} and \cite{Brauer-Lindner-95}, we keep the same names \T{S1}, ..., \T{S14} for the 14 sensors. The reference books is ambiguous with respect to the meaning and role of sensors \T{13} and \T{14}; we resolve this ambiguity by applying the corrigendum described in Appendix~\ref{sec:erratum}.

	To manipulate sensor values, the controller requires only basic types: \T{BOOL}, \T{REAL}, and \T{STRING}. The LNT language provides them as predefined types; this is a clear advantage with respect to LOTOS, in which floating-point numbers and character strings are missing and must be defined explicitly (e.g., by integration of external C code, as it is done in CADP).
	The authors of the simulator mentioned that a precision of $10^{-2}$ is enough when comparing real numbers; this is implemented in the approximate equality function ``\,\T{\~}\,'' defined over real numbers (cf Appendix~\ref{sec:lnt:types}).

	For the internal behaviour of the controller itself, it is convenient to replace these concrete data types by more abstract types having only a few possible values (cf Appendix~\ref{sec:lnt:types}).
	For instance, the Boolean values of the three sensors \T{S1}, \T{S2}, and \T{S3} describing the position of the press are mutually exclusive, because the press cannot be in top, middle, and/or bottom position at the same time; therefore, the position of the press can be better described by a four-valued enumerated type \T{PRESS_POSITION}.

	 Similarly, real numbers can be abstracted away by retaining only their ``significant'' values, i.e., the bounds of the segments in which the controller behaves uniformly.  For instance, to control the elevation of the table, it is sufficient to know whether it is at the lowest level, the highest level, or somewhere in between: we thus abstract the real value of sensor \T{S12} into a three-valued enumerated type \T{TABLE_ANGLE}.

\subsection{The Dispatcher Process}
\label{sec:model:dispatcher}

	The dispatcher (see Appendix~\ref{sec:lnt:dispatcher}) is a cyclical process. In each reaction step, it acquires (using a rendezvous on the gate \T{GET_STATUS}) the concrete values of the 14 sensors and a (possibly empty) list of errors, converts these concrete values into abstract ones (see Section~\ref{sec:model:data}), and dispatches the abstract values to the processes $\T{P}_i$ (using a two-party rendezvous on each gate $\T{G}_i$). For the sake of modularity, the dispatcher only sends to each $\T{P}_i$ the values that are of interest to this process; for instance, process \T{P13}, which controls the motor of the deposit belt, does not receive the current angle of the rotary table. 

	The behaviour of the dispatcher is not necessarily unique, as the abstract values can be sent to the processes $\T{P}_i$ in arbitrary order. Appendix~\ref{sec:lnt:dispatcher} provides two different versions of the dispatcher, one that sends the abstract values in deterministic sequential order (by increasing values of $i$), and another one that sends the abstract values in parallel to all processes $\T{P}_i$.

\subsection{The Individual Processes}
\label{sec:model:individual}

	Each process $\T{P}_i$ (see Appendix~\ref{sec:lnt:controller}) is specified as a parallel composition of two behaviours --- however, the three magnet-related processes \T{P4}, \T{P5}, and \T{P11} contain one single behaviour, whereas process \T{P12} includes a third behaviour that, initially, introduces the five metal blanks into the production cell.

	The first behaviour describes the overall cyclic functioning of a given actuator. Directly derived from the informal specification of the production cell~\cite{Lindner-95}, this behaviour is thus an action loop, possibly preceded by an initial sequence of actions. For instance, process \T{P2}, which controls the cyclic extension and retraction of the first arm of the robot, starts with an initial sequence that brings the robot arm, initially completely retracted, to its minimal value required to start the cycle.

	The second behaviour, which is required for interfacing the asynchronous controller and the synchronous simulator, performs a loop that scrutates the (abstract) sensor values until it is time to issue an actuator command and move to the next state. Both behaviours synchronize on the gate corresponding to this actuator command.

\section{Code Generation from the LOTOS and LNT Specifications}
\label{sec:code-generation}

	Following a ``model-driven'' approach, most of the code of the controller implementation is generated automatically from the LOTOS or LNT specification. This is done using the compilers and the EXEC/C{\AE}SAR software framework~\cite{Garavel-Viho-Zendri-01} provided in the CADP toolbox \cite{Garavel-Lang-Mateescu-Serwe-13}.
	The LOTOS specification is translated to sequential C code (about 7340 lines of C, including blank lines and comments) using the C{\AE}SAR and C{\AE}SARADT compilers of the CADP toolbox.
	The LNT specification is first translated to LOTOS and C code using the LNT2LOTOS compiler, then the generated LOTOS code is translated to C using C{\AE}SAR and C{\AE}SARADT (about 8150 lines of C in total).

	The generated C code is generic, so that it cannot directly connect to the Tcl/Tk simulator. According to the principles of EXEC/C{\AE}SAR, two auxiliary C modules are needed to interface both worlds.

	The first module (750 lines) provides, for each externally visible LOTOS or LNT gate, a corresponding C function. The skeleton of this module can be automatically generated by C{\AE}SAR, so that only the bodies of these gate functions have to be filled in manually. This is straightforward for the functions corresponding to actuator gates (e.g., \T{PRESS_UPWARD}, etc.), as it is sufficient to emit the corresponding simulator command to the standard output.
	The function for gate \T{GET_STATUS} is a bit more complex, as it parses the standard input and converts character strings to LOTOS or LNT values; the most involved parsing task concerns the string containing a list of error messages.

	The second module (90 lines) contains the main function, which explores a (possibly infinite) execution path, following the transitions that are both fireable in the LOTOS or LNT specification and accepted by the Tcl/Tk simulator; if several transitions are possible in a current state, one of them is selected.
	The CADP toolbox provides a standard version of this second module, which in most cases can be used as is. However, in the production cell example, it was necessary to slightly adapt the code in two ways: (i) to send a \T{react} command to the simulator at the end of each reaction step, only after all actuator commands have been emitted, and (ii) to ensure that the \T{get_status} command occurs, and only occurs after a \T{react} command.
	These two constraints express that \T{react} and \T{get_status} have somewhat a lower priority than the commands sent to the actuators; because LOTOS and LNT do not provide priority between transitions, these constraints have been implemented in the main~C program, where the choice between available transitions is actually resolved.

\section{Validation of the LOTOS and LNT Specifications}
\label{sec:validation}

	In this section, we discuss the level of confidence that can be placed in the LOTOS and LNT specifications of the production cell controller.

	First, these specifications passed the stringent compile-time checks performed by the LOTOS and LNT compilers.
	Second, the C code generated from these specifications has been connected to the graphical simulator and intensively exercised, as the simulator continuously provides plausible inputs to the controller (i.e., sensor values respecting the physical constraints of the production cell) and checks the outputs of the controller (i.e., the commands sent to the actuators). Because the simulator signals any error (such as collisions or blanks being dropped) and immediately stops, such a co-simulation is akin to run-time verification.
	We let the controller and the simulator run for five days without observing any problem, which increased our confidence in the correctness of the formal specifications.

	Also, the way our specifications are constructed ensures that certain requirements are satisfied {\em by construction}. For instance, the safety requirements stated in~\cite[Section~2.3.1]{Lindner-95} can be verified by direct inspection of the source specifications:
	(i) the cyclic processes $\T{P}_i$ controlling the actuators clearly keep each movement inside its permitted range, thus avoiding damages caused by out-of-range movements;
	(ii) synchronizing these processes $\T{P}_i$ by multiway rendezvous ensures a coordination of the movements avoiding collisions; for instance, the robot stops its rotation, until the press and the arms are in a position that a further rotation of the robot is safe;
	(iii) similar synchronizations also ensure that blanks are not dropped outside safe areas;
	(iv) each motor is stopped before it is asked to reverse its direction;
	(iv) in each reaction step, at most one command in each actuator group is issued;
	(v) in each reaction step, there is exactly one command \T{react} and one command \T{get_status} issued, etc.

	However, beyond safety properties, \cite[Section~2.3]{Lindner-95} mentions other requirements, such as liveness properties and efficiency, the latter dealing with quantitative time. For such properties, a formal verification would be desirable using, e.g., model checking or equivalence checking, using explicit-state or symbolic state-space exploration, possibly enhanced with partial-order or compositional reduction techniques \cite{Garavel-Lang-Mateescu-15}.
	We have not done this, so we do not know at the moment which approach would be the most suitable for such a challenging task. In the remainder of this section, we simply summarize a few findings from our preliminary attempts.

	A difficulty resides in the process \T{DISPATCHER} added for interfacing the asynchronous controller with the synchronous simulator.
	Indeed, each rendezvous on \T{GET_STATUS} offers all possible values for its fifteen offers (nine Booleans, five reals, and a character string). Even if Boolean combinations are reduced to the admissible ones, even if reals are abstracted to an enumerated type with the twelve essential values used by the sensors (plus another generic value representing all reals different from these twelve ones), and even if the character string is assumed to be constant and ignored, the branching factor for each \T{GET_STATUS} rendezvous would be more than 30,000,000. 
	This suggests to abstract away the controller by removing the \T{DISPATCHER} process and the \T{GET_STATUS} gate. In such an entirely asynchronous model, each process $\T{P}_i$ would receive sensor values directly on its gate $\T{G}_i$, so that a smaller set of real values  (three or four only, see Section~\ref{sec:model:data}) could be associated to each gate $\T{G}_i$. Even then, the branching factor if all gates $\T{G}_i$ are offered simultaneously would be around 20,000.
												
	To make state-space exploration tractable, it seems unavoidable to take into account finer constraints on the sensor values: for instance, the real value of sensor \T{S4} (extension of arm 1) does not evolve randomly, but depends on the commands sent to the corresponding actuator (stable, increasing, or decreasing).
	This would require an accurate modelling of the controller environment, taking inspiration from the graphical simulator code and replicating, in the formal specification, parts of the simulator functionality for generating plausible sensor values.

\section{Conclusion}
\label{sec:conclusion}

	Although the production cell benchmark is now more than twenty-year old, it is still a stimulating example for research in formal methods. This benchmark has several advantages: it is properly described, its requirements are stable and precise and, sadly enough, most of the formal specifications produced for this benchmark in the 90s are no longer available today, which leaves room for the new generation.

	On this case study, we have illustrated the merits of the multiway rendezvous. In an asynchronous concurrency setting, high-level tasks (such as moving the arm of a robot from one point to another in a space with several degrees of freedom) can be simply decomposed into a set of processes that execute simultaneously, most of the time independently, only synchronizing themselves when some goals of common interest have to be reached.
	As there can be more than two such processes, multiway rendezvous is the paradigm of choice to specify an atomic synchronization barrier governing all processes. Multiway rendezvous also supports data communication between these processes, a possibility that was not needed for the production cell, but can be useful to specify, e.g., broadcast or distributed consensus.

	Along these lines, we have shown that multiway rendezvous allows a formal, concise, elegant, and modular description of a software controller for the production cell. Each of the thirteen concurrent processes is responsible for a single operation and can be specified straightforwardly as a cyclic sequence of actions, multiway rendezvous ensuring proper coordination between (subgroups of) these processes. The controller is compositional, in the sense that it can be easily adapted if the architecture of the production cell evolves locally, e.g., by adding new devices or removing existing ones --- this is the {\em flexibility\/} requirement mentioned in \cite[Section~2.3.3]{Lindner-95}.

	The software controller was successively specified in LOTOS, then in LNT. The complete LNT specification, which is more readable than the LOTOS one, is provided in Appendix~\ref{sec:lnt}. For both specifications, the CADP toolbox generated an implementation in C that was connected, using the EXEC/C{\AE}SAR interface, to the Tcl/Tk simulator and used to drive the production cell.

	Although the LOTOS and LNT specifications pass the compile-time checks of the CADP compiler and the run-time checks of the Tcl/Tk simulator, they have not been yet formally verified using, e.g., model checking or equivalence checking. Their verification thus remains a challenging problem for future work.

\subsubsection*{Acknowledgements}

	We are grateful to Artur Brauer, Thomas Lindner, and Claus Lewerentz for valuable discussions and information about the case study and the graphical simulator, and to Mark Jorgensen, who contributed to the third version (1997) of the LOTOS specification.

\appendix

\clearpage

\section{Overview of Related Work}
\label{sec:related-work}

	As mentioned in Section~\ref{sec:prior-work}, there have been numerous applications of formal methods to the production cell case study --- at least thirty, including our LOTOS and LNT specifications.

	We present them in two tables. Table~\ref{tab:descriptive-models} lists the ``descriptive'' approaches, in which formal methods have been used only for specification or verification purposes. Table~\ref{tab:connected-models} gathers the approaches in which, as reported in the corresponding published articles, an executable controller was automatically derived from the formal specification and connected to the graphical simulator.

	Each table indicates whether executable code was generated automatically, manually, or by refinement (column 2), whether the specification was connected to the simulator (column 3), whether the formal specification uses multiway rendezvous (column 4), and, if available, the size of the specification (column 5).

\begin{table}[h]
  \caption{Descriptive approaches}
  \label{tab:descriptive-models}  
  \small
  \begin{center}
    \begin{tabular}{lcccc}
      \hline
      language/tool & code generation & simulation & multiway & size \\
      \hline
      ASM \cite{Boerger-Mearelli-97} & refinement & yes & no & 9 pages (ground model) \\
      CO-OPN \cite{Barbey-Buchs-Pereire-98} & yes & no & no & 100 pages \\
      Coordinated Atomic Actions \cite{Zorzo-Romanovsky-Xu-et-al-99,Xu-Randell-Romanovsky-et-al-02} & manual & yes & no & 4500 lines \\
      CSP \& CML \cite{Rischel-Sun-97} & manual & yes & no & \\
      CTA \cite{Beyer-Rust-98} \cite{Rust-99} \cite{Beyer-02} & & no & no & \\
      Duration Calculus \cite{Aminuddin-Jifeng-Abdullah-01} \cite{Fraenzle-96} & possible & no & no & \\
      Event-B \cite{Gondal-12} \cite{Gondal-Poppleton-Butler-11} & refinement & no & no & \\
      Focus \cite{Fuchs-Philipps-95} & no & no & no & 80 lines \\
      HOL \cite{CardellOliver-95} & no & no & no & 650 lines \\
      KIV \cite{Schellhorn-Burandt-95} & no & no & no & 2000 lines, 611 axioms \\
      LCM \& MCM \cite{Wieringa-95} & no & no & no & 8 pages \\
      Modula-3 \cite{Rueping-Sekerinski-95} & yes & no & no & 1400 lines \\
      PEPA \cite{Holton-95} & no & no & no & \\
      RAISE \cite{Erasmy-Sekerinksi-94} \cite{Erasmy-Sekerinski-95} & no & no & no & 676 lines \\
      SDL \cite{Heinkel-Lindner-95} & yes & not tried & no & 1800 lines \\
      Spectrum \cite{Dranidis-Gastinger-95} & no & no & no & \\
      Statecharts \cite{Damm-Hungar-Kelb-et-al-95} & no & no & no & 8.4E+19 states \\
      SYSYPHOS \cite{Burghardt-95} \cite{Burghardt-96} & circuit & no & no & incomplete \\
      Tatzelwurm \cite{Klingenbeck-Kaeufl-95} & no & no & no & incomplete \\
      TLT \cite{Cuellar-Huber-95} & no & manual & no & \\
      Troll-\emph{light} \cite{Herzig-Vlachantonis-95} & no & no & no & incomplete \\
      UML \cite{Lilius-Paltor-00} & no & no & no & 52,060 states \\
      Z \& Petri Nets \cite{Heiner-Deussen-95} \cite{Heiner-Deussen-Spranger-96} \cite{Heiner-Heisel-99} & no & no & no & 51 places, 36 transitions\\
      \hline
    \end{tabular}
  \end{center}
\end{table}

\begin{table}
  \caption{Executable approaches}
  \label{tab:connected-models}
  
  \begin{center}
    \small
    \begin{tabular}{lcccc}
      \hline
      language/tool & code generation & simulation & multiway & size \\
      \hline
      CSL \cite{Melcher-Winkelmann-98} \cite{Noekel-Winkelmann-95} & yes & yes & no & 9 pages \\
      Esterel \cite{Budde-95} & yes & yes & no & 400 lines \\
      \textbf{LNT} & \textbf{yes} & \textbf{yes} & \textbf{yes} & \textbf{804 lines} \\
      \textbf{LOTOS} & \textbf{yes} & \textbf{yes} & \textbf{yes} & \textbf{753 lines} \\
      Lustre \cite{Holenderski-95} & yes & yes & no & 200 lines \\
      Signal \cite{Amagbegnon-LeGuernic-Marchand-et-al-95-b} \cite{Amagbegnon-LeGuernic-Marchand-et-al-95-a} & yes & yes & no & 1700 lines \\
      STD \& ICOS${}^2$ \cite{Korf-Schloer-95} & yes & yes & no & 62 timing diagrams \\
      \hline
    \end{tabular}
  \end{center}
\end{table}

\section{Errata in the Task Description}
\label{sec:erratum}

	The reader interested in the production cell might have noticed an inconsistency between the task description~\cite{Lindner-95} and the specification of the graphical simulator~\cite{Brauer-Lindner-95}:
	Chapter~2 of the reference book \cite[Section~2.1 (page~15) and Section~2.3.1 (page~18)]{Lindner-95} states that sensor~13 is associated to the deposit belt, and sensor~14 to the feed belt, whereas Appendix~A of the same book \cite[Table~1, page~390]{Brauer-Lindner-95} states exactly the opposite.
	After discussion with the authors, it appears that Appendix~A is right.
Thus, the following changes should be applied to Chapter~2 of the reference book \cite{Lindner-95}:

\begin{itemize}
	\item On page~15, the items 13 and 14 of the enumeration should be permuted.

	\item On page~15, the second to last paragraph of Section~2.2.2 should be modified as follows: {\em Both photoelectric cells switch on when a plate intercepts the light ray. Just after the plate has completely passed through it, the light barrier switches off. At this precise moment, the plate is in the correct position to be picked up by the travelling crane (sensor 14 of the deposit belt), respectively it has just left the belt to land on the elevating rotary table --- provided of course that the latter machine is correctly positioned --- (sensor 13 of the feed belt)}.

	\item On page~18, the first item of the section entitled ``Keep blanks sufficiently distant'' should be modified as follows: {\em a new blank may only be put on the feed belt, if sensor 13 confirms that the last one has arrived at the end of the feed belt}.

	\item On page~18, the second item of the same section should be modified as follows: {\em a new blank may only be put on the deposit belt, if sensor 14 confirms that the last one has arrived at the end of the deposit belt}.
\end{itemize}

\clearpage

\section{LNT Specification of the Production Cell Controller}
\label{sec:lnt}

Our LNT specification of the production cell controller is decomposed in six modules.

\subsection{Module TYPES}
\label{sec:lnt:types}

This module implements the data abstractions presented in Sect.~\ref{sec:model:data}. It defines: (i) the approximate equality function ``\,\T{\~}\,'' defined over real numbers; (ii) the enumerated types that abstract sensor values; (iii) the conversion functions to convert (tuples of) sensor values into abstract values of these enumerated types.

\begin{lstlisting}
module TYPES with "==" is

function _~_ (X1, X2: REAL) : BOOL is
   -- equality of two reals up to precision 10^-2
   return (abs (X1 - X2) < 1.0e-2)
end function

--------------------------------------------------------

type PRESS_POSITION is
   -- position of the press
   PRESS_BOTTOM,
   PRESS_MIDDLE,
   PRESS_TOP,
   OTHER
end type

--------------------------------------------------------

type ARM1_EXTENSION is
   -- extension of arm 1
   ARM1_MIN, -- 0.5208
   ARM1_MAX, -- 0.6458
   OTHER
end type

--------------------------------------------------------

type ARM2_EXTENSION is
   -- extension of arm 2
   ARM2_MAX, -- 0.7971
   ARM2_MIN, -- 0.5707
   OTHER
end type

--------------------------------------------------------

type ROBOT_ANGLE is
   -- angle of the robot
   ROBOT_M90, -- -90
   ROBOT_M70, -- -70
   ROBOT_0,
   ROBOT_15,
   ROBOT_35,
   ROBOT_50,
   OTHER
end type

--------------------------------------------------------

type TABLE_POSITION is
   -- position of the table
   TABLE_BOTTOM,
   TABLE_TOP,
   OTHER
end type

--------------------------------------------------------

type TABLE_ANGLE is
   -- angle of the table
   ANGLE_MIN, -- 0
   ANGLE_MAX, -- 50
   OTHER
end type

--------------------------------------------------------

type CRANE_POSITION is
   -- position of the crane
   CRANE_OVER_FEED_BELT,
   CRANE_OVER_DEPOSIT_BELT,
   OTHER
end type

--------------------------------------------------------

type CRANE_HEIGHT is
   -- height of the crane
   CRANE_HIGH, -- 0.9450
   CRANE_LOW,  -- 0.6593
   OTHER
end type

--------------------------------------------------------

function CONVERT_S1_S2_S3 (S1, S2, S3: BOOL) : PRESS_POSITION is
   -- this function converts signals received from the Tcl/Tk simulator
   -- into the corresponding abstract values representing the position
   -- of the press
   assert not (S1 and S2) and not (S1 and S3) and not (S2 and S3);
   -- at most one of S1, S2, and S3 is true
   if S1 then
      return PRESS_BOTTOM
   elsif S2 then
      return PRESS_MIDDLE
   elsif S3 then
      return PRESS_TOP
   else
      return OTHER
   end if
end function

--------------------------------------------------------

function CONVERT_S4 (S4: REAL) : ARM1_EXTENSION is
   -- this function converts signals received from the Tcl/Tk simulator
   -- into the corresponding abstract values representing the extension
   -- of arm 1
   if S4 ~ 0.5208 then
      return ARM1_MIN
   elsif S4 ~ 0.6458 then
      return ARM1_MAX
   else
      return OTHER
   end if
end function

--------------------------------------------------------

function CONVERT_S5 (S5: REAL) : ARM2_EXTENSION is
   -- this function converts signals received from the Tcl/Tk simulator
   -- into the corresponding abstract values representing the extension
   -- of arm 2
   if S5 ~ 0.7971 then
      return ARM2_MAX
   elsif S5 ~ 0.5707 then
      return ARM2_MIN
   else
      return OTHER
   end if
end function

--------------------------------------------------------

function CONVERT_S6 (S6: REAL) : ROBOT_ANGLE is
   -- this function converts signals received from the Tcl/Tk simulator
   -- into the corresponding abstract values representing the height of
   -- the robot's angle
   if S6 ~ -90.0 then
      return ROBOT_M90
   elsif S6 ~ -70.0 then
      return ROBOT_M70
   elsif S6 ~ 0.0 then
      return ROBOT_0
   elsif S6 ~ 15.0 then
      return ROBOT_15
   elsif S6 ~ 35.0 then
      return ROBOT_35
   elsif S6 ~ 50.0 then
      return ROBOT_50
   else
      return OTHER
   end if
end function

--------------------------------------------------------

function CONVERT_S7_S8 (S7, S8: BOOL) : TABLE_POSITION is
   -- this function converts signals received from the Tcl/Tk simulator
   -- into the corresponding abstract values representing the position
   -- of the table
   assert not (S7 and S8); -- at most one of S7 and S8 is true
   if S7 then
      return TABLE_BOTTOM
   elsif S8 then
      return TABLE_TOP
   else
      return OTHER
   end if
end function

--------------------------------------------------------

function CONVERT_S9 (S9: REAL) : TABLE_ANGLE is
   -- this function converts signals received from the Tcl/Tk simulator
   -- into the corresponding abstract values representing the angle of
   -- the table
   if S9 ~ 0.0 then
      return ANGLE_MIN
   elsif S9 ~ 50.0 then
      return ANGLE_MAX
   else
      return OTHER
   end if
end function

--------------------------------------------------------

function CONVERT_S10_S11 (S10, S11: BOOL) : CRANE_POSITION is
   -- this function converts signals received from the Tcl/Tk simulator
   -- into the corresponding abstract values representing the positions
   -- of the crane
   assert not (S10 and S11); -- at most one of S10 and S11 is true
   if S10 then
      return CRANE_OVER_DEPOSIT_BELT
   elsif S11 then
      return CRANE_OVER_FEED_BELT
   else
      return OTHER
   end if
end function

--------------------------------------------------------

function CONVERT_S12 (S12: REAL) : CRANE_HEIGHT is
   -- this function converts signals received from the Tcl/Tk simulator
   -- into the corresponding abstract values representing the height of
   -- the crane
   if S12 ~ 0.9450 then
      return CRANE_HIGH
   elsif S12 ~ 0.6593 then
      return CRANE_LOW
   else
      return OTHER
   end if
end function

end module
\end{lstlisting}

\subsection{Module STATES}
\label{sec:lnt:states}

This module defines: (i) three enumerated types that encode the states of cyclic behaviours of individual processes; (ii) three next-state functions \T{SUCC} for these types; (iii) eight functions \T{LIMIT_}$xxx$ that express when an device of the production cell has reached a point where it must proceed to its next state.

\begin{lstlisting}
module STATES (TYPES) with "==" is

type TWO_STATE is
   -- this type is used in processes P2, P3, P7, P8, P9, and P10
   1, 2
end type

function SUCC (S: TWO_STATE) : TWO_STATE is
   case S in
      1 -> return 2
    | 2 -> return 1
   end case
end function

--------------------------------------------------------

type THREE_STATE is
   -- this type is used in process P1
   1, 2, 3
end type

function SUCC (S: THREE_STATE) : THREE_STATE is
   case S in
      1 -> return 2
    | 2 -> return 3
    | 3 -> return 1
   end case
end function

--------------------------------------------------------

type ELEVEN_STATE is
   -- this type is used in process P6
   i1, i2, i3, i4, 1, 2, 3, 4, 5, 6, 7
end type

function SUCC (S: ELEVEN_STATE) : ELEVEN_STATE is
   case S in
      i1 -> return i2
    | i2 -> return i3
    | i3 -> return i4
    | i4 -> return  1
    |  1 -> return  2
    |  2 -> return  3
    |  3 -> return  4
    |  4 -> return  5
    |  5 -> return  6
    |  6 -> return  7
    |  7 -> return  1
   end case
end function

--------------------------------------------------------

-- functions that return "true" when a particular engine has reached
-- a specified limit of movement, so that a state change is required

--------------------------------------------------------

function LIMIT_PRESS_POSITION (STATE: THREE_STATE, VALUE: PRESS_POSITION) : BOOL is
   return ((STATE == 1) and (VALUE == PRESS_BOTTOM)) or
          ((STATE == 2) and (VALUE == PRESS_MIDDLE)) or
          ((STATE == 3) and (VALUE == PRESS_TOP))
end function

--------------------------------------------------------

function LIMIT_ARM1_EXTENSION (STATE: TWO_STATE, VALUE: ARM1_EXTENSION) : BOOL is
   return ((STATE == 1) and (VALUE == ARM1_MIN)) or
          ((STATE == 2) and (VALUE == ARM1_MAX))
end function

--------------------------------------------------------

function LIMIT_ARM2_EXTENSION (STATE: TWO_STATE, VALUE: ARM2_EXTENSION) : BOOL is
   return ((STATE == 1) and (VALUE == ARM2_MAX)) or
          ((STATE == 2) and (VALUE == ARM2_MIN))
end function

--------------------------------------------------------

function LIMIT_ROBOT_ANGLE (STATE: ELEVEN_STATE, VALUE: ROBOT_ANGLE) : BOOL is
   return ((STATE == i1) and (VALUE == ROBOT_15))  or
          ((STATE == i2) and (VALUE == ROBOT_50))  or
          ((STATE == i3) and (VALUE == ROBOT_15))  or
          ((STATE == i4) and (VALUE == ROBOT_M70)) or
          ((STATE ==  1) and (VALUE == ROBOT_M90)) or
          ((STATE ==  2) and (VALUE == ROBOT_M70)) or
          ((STATE ==  3) and (VALUE == ROBOT_15))  or
          ((STATE ==  4) and (VALUE == ROBOT_50))  or
          ((STATE ==  5) and (VALUE == ROBOT_35))  or
          ((STATE ==  6) and (VALUE == ROBOT_15))  or
          ((STATE ==  7) and (VALUE == ROBOT_M70))
end function

--------------------------------------------------------

function LIMIT_TABLE_POSITION (STATE: TWO_STATE, VALUE: TABLE_POSITION) : BOOL is
   return ((STATE == 1) and (VALUE == TABLE_BOTTOM)) or
          ((STATE == 2) and (VALUE == TABLE_TOP))
end function

--------------------------------------------------------

function LIMIT_TABLE_ANGLE (STATE: TWO_STATE, VALUE: TABLE_ANGLE) : BOOL is
   return ((STATE == 1) and (VALUE == ANGLE_MIN)) or
          ((STATE == 2) and (VALUE == ANGLE_MAX))
end function

--------------------------------------------------------

function LIMIT_CRANE_POSITION (STATE: TWO_STATE, VALUE: CRANE_POSITION) : BOOL is
   return ((STATE == 1) and (VALUE == CRANE_OVER_DEPOSIT_BELT)) or
          ((STATE == 2) and (VALUE == CRANE_OVER_FEED_BELT))
end function

--------------------------------------------------------

function LIMIT_CRANE_HEIGHT (STATE: TWO_STATE, VALUE: CRANE_HEIGHT) : BOOL is
   return ((STATE == 1) and (VALUE == CRANE_HIGH)) or
          ((STATE == 2) and (VALUE == CRANE_LOW))
end function

end module
\end{lstlisting}

\subsection{Module CHANNELS}
\label{sec:lnt:channels}

This module defines the channel types for the gate \T{GET_STATUS}, which transports concrete sensor values, and for the gates $\T{G}_i$, which transport abstract sensor values.

\begin{lstlisting}
module CHANNELS (TYPES) is

-- definition of the channel types, many of them are overloaded with the
-- name of the corresponding type

channel STATUS is
   -- simulation status: values of all sensors
   (BOOL,        -- S1:  press in bottom position
    BOOL,        -- S2:  press in middle position
    BOOL,        -- S3:  press in top position
    REAL,        -- S4:  extension of the robot's arm 1
    REAL,        -- S5:  extension of the robot's arm 2
    REAL,        -- S6:  angle of rotation of the robot
    BOOL,        -- S7:  elevating rotary table in bottom position
    BOOL,        -- S8:  elevating rotary table in top position
    REAL,        -- S9:  angle of rotation of the table
    BOOL,        -- S10: crane over the deposit belt
    BOOL,        -- S11: crane over the feed belt
    REAL,        -- S12: height of the crane's magnet
    BOOL,        -- S13: blank inside the feed belt photoelectric barrier
    BOOL,        -- S14: blank inside the deposit belt photoelectric barrier
    STRING       -- E:   errors that occurred since the last reaction cycle
    )
end channel

channel BOOL is
   (BOOL)
end channel

channel PRESS_POSITION is
   (PRESS_POSITION)
end channel

channel ARM1_EXTENSION is
   (ARM1_EXTENSION)
end channel

channel ARM2_EXTENSION is
   (ARM2_EXTENSION)
end channel

channel ROBOT_ANGLE is
   (ROBOT_ANGLE)
end channel

channel TABLE_ANGLE is
   (TABLE_ANGLE)
end channel

channel TABLE_POSITION is
   (TABLE_POSITION)
end channel

channel CRANE_POSITION is
   (CRANE_POSITION)
end channel

channel CRANE_HEIGHT is
   (CRANE_HEIGHT)
end channel

end module
\end{lstlisting}

\subsection{Module DISPATCHER}
\label{sec:lnt:dispatcher}

This module defines the \T{DISPATCHER} process described in Sect.~\ref{sec:model:dispatcher}. Depending on the value of the Boolean parameter \T{SEQUENTIAL}, abstract values will be sent sequentially or concurrently to the gates $\T{G}_i$. 

\begin{lstlisting}
module DISPATCHER (TYPES, CHANNELS) is

process DISPATCHER [GET_STATUS: STATUS,
                    G1: PRESS_POSITION,
                    G2: ARM1_EXTENSION,
                    G3: ARM2_EXTENSION,
                    G6: ROBOT_ANGLE,
                    G7: TABLE_POSITION,
                    G8: TABLE_ANGLE,
                    G9: CRANE_POSITION,
                    G10: CRANE_HEIGHT,
                    G12, G13: BOOL]
                    (SEQUENTIAL: BOOL) is
   -- this process receives inputs from the Tcl/Tk simulator and dispatches
   -- them to the corresponding components of the controller process
   var
      S1, S2, S3, S7, S8, S10, S11, S13, S14: BOOL,
      S4, S5, S6, S9, S12: REAL
   in
      loop
         GET_STATUS (?S1, ?S2, ?S3, ?S4, ?S5, ?S6, ?S7, ?S8, ?S9, ?S10, ?S11,
                     ?S12, ?S13, ?S14, ?any STRING);
         if SEQUENTIAL then
            -- inputs are dispatched to controller gates in sequential order,
            -- which reduces the amount of nondeterminism, possibly making the 
            -- specification easier to analyze
            G1 (CONVERT_S1_S2_S3 (S1, S2, S3));      -- press position
            G2 (CONVERT_S4 (S4));                    -- arm1 extension
            G3 (CONVERT_S5 (S5));                    -- arm2 extension
            G6 (CONVERT_S6 (S6));                    -- robot angle
            G7 (CONVERT_S7_S8 (S7, S8));             -- table position
            G8 (CONVERT_S9 (S9));                    -- table angle
            G9 (CONVERT_S10_S11 (S10, S11));         -- crane position
            G10 (CONVERT_S12 (S12));                 -- crane height
            G12 (S13);                               -- sensor feed belt
            G13 (S14)                                -- sensor deposit belt
         else
            -- inputs are dispatched to controller gates in any order
            par
               G1 (CONVERT_S1_S2_S3 (S1, S2, S3))    -- press position
            || G2 (CONVERT_S4 (S4))                  -- arm1 extension
            || G3 (CONVERT_S5 (S5))                  -- arm2 extension
            || G6 (CONVERT_S6 (S6))                  -- robot angle
            || G7 (CONVERT_S7_S8 (S7, S8))           -- table position
            || G8 (CONVERT_S9 (S9))                  -- table angle
            || G9 (CONVERT_S10_S11 (S10, S11))       -- crane position
            || G10 (CONVERT_S12 (S12))               -- crane height
            || G12 (S13)                             -- sensor feed belt
            || G13 (S14)                             -- sensor deposit belt
            end par
         end if
      end loop
   end var
end process

end module
\end{lstlisting}

\subsection{Module CONTROLLER}
\label{sec:lnt:controller}

This module defines a process named \T{CONTROLLER} that achieves the parallel composition of the individual processes described in Sect.~\ref{sec:model:individual}, and then these thirteen individual processes \T{P1}, ... \T{P13} themselves.
	The controller process handles three sets of gates: (i) the gates $\T{G}_i$ of the processes $\T{P}_i$; (ii) the external gates (\T{PRESS_UPWARD}, ...) used to send actuator commands to the graphical simulator; and (iii) the internal gates (\T{FT_READY}, ...) used to synchronize the $\T{P}_i$ processes using binary or multiway rendezvous.
	 Notice that the ``graphical'' $n$-ary parallel composition of LNT~\cite{Garavel-Sighireanu-99} allows to represent the controller process concisely, rather than breaking it into many binary parallel operators with involved synchronization sets.

\begin{lstlisting}
module CONTROLLER (TYPES, CHANNELS, STATES) is

process CONTROLLER [G1: PRESS_POSITION,
                    G2: ARM1_EXTENSION,
                    G3: ARM2_EXTENSION,
                    G6: ROBOT_ANGLE,
                    G7: TABLE_POSITION,
                    G8: TABLE_ANGLE,
                    G9: CRANE_POSITION,
                    G10: CRANE_HEIGHT,
                    G12, G13: BOOL,
                    PRESS_UPWARD, PRESS_STOP, PRESS_DOWNWARD,
                    ARM1_FORWARD, ARM1_STOP, ARM1_BACKWARD,
                    ARM2_FORWARD, ARM2_STOP, ARM2_BACKWARD,
                    ARM1_MAG_ON, ARM1_MAG_OFF,
                    ARM2_MAG_ON, ARM2_MAG_OFF,
                    ROBOT_LEFT, ROBOT_STOP, ROBOT_RIGHT,
                    TABLE_LEFT, TABLE_STOP_H, TABLE_RIGHT,
                    TABLE_UPWARD, TABLE_STOP_V, TABLE_DOWNWARD,
                    CRANE_TO_BELT2, CRANE_STOP_H, CRANE_TO_BELT1,
                    CRANE_LIFT, CRANE_STOP_V, CRANE_LOWER,
                    CRANE_MAG_ON, CRANE_MAG_OFF,
                    BELT1_START, BELT1_STOP,
                    BELT2_START, BELT2_STOP,
                    BLANK_ADD: NONE] is

   -- the controller consists in 13 concurrent processes P1, ..., P13, each
   -- supervising a particular engine of the production cell, or a given
   -- degree of freedom of a particular engine

   hide
      -- each gate is noted [3], [4], or [5] if it is used in three-party,
      -- four-party, or five-party rendezvous, respectively; absence of such
      -- indication means that the gate is used in two-party rendezvous
      FT_READY,      -- belt1 ready to deliver a blank element to the table [3]
      FT,            -- belt1 delivers a blank element to the table [3]
      TA1_READY,     -- table ready to deliver a blank element to arm1 [5]
      TA1,           -- arm1 took a blank element from the table [3]
      A1P,           -- arm1 ready to deliver a blank element to the press [4]
      PA2,           -- press delivers an element to arm2 [4]
      A2D,           -- arm2 puts a pressed element on belt2 [4]
      DC_READY,      -- crane arrived over belt2
      DC,            -- crane gets a pressed element from belt2 [4]
      CF,            -- crane puts a blank element on belt1 [4]
      UP_M70,        -- robot angle is or will soon be greater than -70 degr.
      UP_15,         -- robot angle is or will soon be greater than 15 degr.
      DOWN_15,       -- robot angle is or will soon be smaller than 15 degr.
      DOWN_M70: NONE -- robot angle is or will soon be smaller than -70 degr.
   in
      par
         A1P, PA2, UP_M70, UP_15, DOWN_15, DOWN_M70 ->
            P1 [G1, PRESS_UPWARD, PRESS_STOP, PRESS_DOWNWARD,
                A1P, PA2, UP_M70, UP_15, DOWN_15, DOWN_M70]
       ||
         TA1_READY, A1P ->
            P2 [G2, ARM1_FORWARD, ARM1_STOP, ARM1_BACKWARD, TA1_READY, A1P]
       ||
         PA2, A2D ->
            P3 [G3, ARM2_FORWARD, ARM2_STOP, ARM2_BACKWARD, PA2, A2D]
       ||
         TA1_READY, A1P, TA1 ->
            P4 [ARM1_MAG_ON, ARM1_MAG_OFF, TA1_READY, A1P, TA1]
       ||
         PA2, A2D ->
            P5 [ARM2_MAG_ON, ARM2_MAG_OFF, PA2, A2D]
       ||
         TA1_READY, A1P, PA2, A2D, UP_M70, UP_15, DOWN_15, DOWN_M70 ->
            P6 [G6, ROBOT_LEFT, ROBOT_STOP, ROBOT_RIGHT,
                TA1_READY, A1P, PA2, A2D, UP_M70, UP_15, DOWN_15, DOWN_M70]
       ||
         FT_READY, FT, TA1_READY, TA1 ->
            P7 [G7, TABLE_UPWARD, TABLE_STOP_V, TABLE_DOWNWARD,
                FT_READY, FT, TA1_READY, TA1]
       ||
         FT_READY, FT, TA1_READY, TA1 ->
            P8 [G8, TABLE_LEFT, TABLE_STOP_H, TABLE_RIGHT,
                FT_READY, FT, TA1_READY, TA1]
       ||
         DC_READY, DC, CF ->
            P9 [G9, CRANE_TO_BELT2, CRANE_STOP_H, CRANE_TO_BELT1,
                DC_READY, DC, CF]
       ||
         DC_READY, DC, CF ->
            P10 [G10, CRANE_LIFT, CRANE_STOP_V, CRANE_LOWER, DC_READY, DC, CF]
       ||
         DC, CF ->
            P11 [CRANE_MAG_ON, CRANE_MAG_OFF, DC, CF]
       ||
         FT_READY, FT, CF ->
            P12 [G12, BELT1_START, BELT1_STOP, BLANK_ADD, FT_READY, FT, CF]
       ||
         A2D, DC ->
            P13 [G13, BELT2_START, BELT2_STOP, A2D, DC]
      end par
   end hide
end process

--------------------------------------------------------

-- each process Pi is split into two (but sometimes one, or sometimes three)
-- concurrent processes; the former process describes the overall functioning
-- cycle of the engine, while the latter process scrutates the inputs and
-- decides when a transition to a next state is required

--------------------------------------------------------

process P1 [G1: PRESS_POSITION,
            PRESS_UPWARD, PRESS_STOP, PRESS_DOWNWARD,
            A1P, PA2, UP_M70, UP_15, DOWN_15, DOWN_M70: NONE] is
   -- this process controls the press
   -- initially, the press is in middle position
   par PRESS_STOP in
      -- the actions before the loop are the same as the actions inside the
      -- loop starting from state 1, but without the rendezvous on gate PA2;
      -- indeed, initially, there is no item in the press that could be
      -- delivered to arm 2
      PRESS_DOWNWARD;
      PRESS_STOP; -- bottom position -> state 2
      UP_15;
      loop
         DOWN_15;
         PRESS_UPWARD;
         PRESS_STOP; -- middle position -> state 3
         DOWN_M70;
         A1P;
         UP_M70;
         PRESS_UPWARD;
         PRESS_STOP; -- top position -> state 1
         PRESS_DOWNWARD;
         PRESS_STOP; -- bottom position -> state 2
         UP_15;
         PA2
      end loop
    ||
      var
         STATE: THREE_STATE,
         VALUE: PRESS_POSITION
      in
         STATE := 1;
         loop
            G1 (?VALUE);
            if LIMIT_PRESS_POSITION (STATE, VALUE) then
               PRESS_STOP;
               STATE := SUCC (STATE)
            end if
         end loop
      end var
   end par
end process

--------------------------------------------------------

process P2 [G2: ARM1_EXTENSION,
            ARM1_FORWARD, ARM1_STOP, ARM1_BACKWARD, TA1_READY, A1P: NONE] is
   -- this process controls the extension of arm 1
   -- initially, arm 1 is completely retracted
   par ARM1_STOP in
      -- arm 1 is initially out of its range of operation (it is too short);
      -- thus, it must first be extended to the lower limit of its range of
      -- operation
      ARM1_FORWARD;
      loop
         ARM1_STOP; -- 0.5208
         TA1_READY;
         ARM1_FORWARD;
         ARM1_STOP; -- 0.6458
         A1P;
         ARM1_BACKWARD
      end loop
    ||
      var STATE: TWO_STATE,
          VALUE: ARM1_EXTENSION
      in
         STATE := 1;
         loop
            G2 (?VALUE);
            if LIMIT_ARM1_EXTENSION (STATE, VALUE) then
               ARM1_STOP;
               STATE := SUCC (STATE)
            end if
         end loop
      end var
   end par
end process

--------------------------------------------------------

process P3 [G3: ARM2_EXTENSION,
            ARM2_FORWARD, ARM2_STOP, ARM2_BACKWARD, PA2, A2D: NONE] is
   -- this process controls the extension of arm 2
   -- initially, arm 2 is completely retracted
   par ARM2_STOP in
      loop
         ARM2_FORWARD;
         ARM2_STOP; -- 0.7971
         PA2;
         ARM2_BACKWARD;
         ARM2_STOP; -- 0.5707
         A2D
      end loop
    ||
      var STATE: TWO_STATE,
          VALUE: ARM2_EXTENSION
      in
         STATE := 1;
         loop
            G3 (?VALUE);
            if LIMIT_ARM2_EXTENSION (STATE, VALUE) then
               ARM2_STOP;
               STATE := SUCC (STATE)
            end if
         end loop
      end var
   end par
end process

--------------------------------------------------------

process P4 [ARM1_MAG_ON, ARM1_MAG_OFF, TA1_READY, A1P, TA1: NONE] is
   -- this process controls the magnet of arm 1
   -- initially, the magnet of arm 1 is off
   loop
      TA1_READY;
      ARM1_MAG_ON;
      TA1;
      A1P;
      ARM1_MAG_OFF
   end loop
end process

--------------------------------------------------------

process P5 [ARM2_MAG_ON, ARM2_MAG_OFF, PA2, A2D: NONE] is
   -- this process controls the magnet of arm 2
   -- initially, the magnet of arm 2 is off
   loop
      PA2;
      ARM2_MAG_ON;
      A2D;
      ARM2_MAG_OFF
   end loop
end process

--------------------------------------------------------

process P6 [G6: ROBOT_ANGLE,
            ROBOT_LEFT, ROBOT_STOP, ROBOT_RIGHT,
            TA1_READY, A1P, PA2, A2D,
            UP_M70, UP_15, DOWN_15, DOWN_M70: NONE] is
   -- this process controls the angle of the robot
   -- initially, the angle of the robot is 0 degrees
   par ROBOT_STOP in
      -- the actions before the loop are the same as the actions inside the
      -- loop starting from state 4, but without the stop at 35 degrees,
      -- the rendezvous on PA2, and the restart of the movement to the left;
      -- indeed, initially there is no item in the press that could be
      -- delivered to arm 2
      ROBOT_RIGHT;
      ROBOT_STOP; -- 15 degrees -> state i2
      UP_15;
      ROBOT_RIGHT;
      ROBOT_STOP; -- 50 degrees -> state i3
      TA1_READY;
      ROBOT_LEFT;
      ROBOT_STOP; -- 15 degrees -> state i4
      DOWN_15;
      ROBOT_LEFT;
      ROBOT_STOP; -- -70 degrees -> state 1
      DOWN_M70;
      loop
         ROBOT_LEFT;
         ROBOT_STOP; -- -90 degrees -> state 2
         A1P;
         ROBOT_RIGHT;
         ROBOT_STOP; -- -70 degrees -> state 3
         UP_M70;
         ROBOT_RIGHT;
         ROBOT_STOP; -- 15 degrees -> state 4
         UP_15;
         ROBOT_RIGHT;
         ROBOT_STOP; -- 50 degrees -> state 5
         TA1_READY;
         ROBOT_LEFT;
         ROBOT_STOP; -- 35 degrees -> state 6
         PA2;
         ROBOT_LEFT;
         ROBOT_STOP; -- 15 degrees -> state 7
         DOWN_15;
         ROBOT_LEFT;
         ROBOT_STOP; -- -70 degrees -> state 1
         DOWN_M70;
         A2D
      end loop
    ||
      var STATE: ELEVEN_STATE,
          VALUE: ROBOT_ANGLE
      in
         STATE := i1;
         loop
            G6 (?VALUE);
	    if LIMIT_ROBOT_ANGLE (STATE, VALUE) then
               ROBOT_STOP;
               STATE := SUCC (STATE)
            end if
         end loop
      end var
   end par
end process

--------------------------------------------------------

process P7 [G7: TABLE_POSITION,
            TABLE_UPWARD, TABLE_STOP_V, TABLE_DOWNWARD,
            FT_READY, FT, TA1_READY, TA1: NONE] is
   -- this process controls the height of the table
   -- initially, the table is in bottom position
   par TABLE_STOP_V in
      TABLE_STOP_V; -- initialisation is mandatory
      loop
         FT_READY;
         FT;
         TABLE_UPWARD;
         TABLE_STOP_V;
         TA1_READY;
         TA1;
         TABLE_DOWNWARD;
         TABLE_STOP_V
      end loop
    ||
      var STATE: TWO_STATE,
          VALUE: TABLE_POSITION -- initial value is TABLE_BOTTOM
      in
         STATE := 1;
         loop
            G7 (?VALUE);
            if LIMIT_TABLE_POSITION (STATE, VALUE) then
               TABLE_STOP_V;
               STATE := SUCC (STATE)
            end if
         end loop
      end var
   end par
end process

--------------------------------------------------------

process P8 [G8: TABLE_ANGLE,
            TABLE_LEFT, TABLE_STOP_H, TABLE_RIGHT,
            FT_READY, FT, TA1_READY, TA1: NONE] is
   -- this process controls the angle of the table
   -- initially, the angle of the table is 0 degrees
   par TABLE_STOP_H in
      TABLE_STOP_H; -- initialisation is mandatory
      loop
         FT_READY;
         FT;
         TABLE_RIGHT;
         TABLE_STOP_H;
         TA1_READY;
         TA1;
         TABLE_LEFT;
         TABLE_STOP_H
      end loop
    ||
      var STATE: TWO_STATE,
          VALUE: TABLE_ANGLE
      in
         STATE := 1;
         loop
            G8 (?VALUE);  -- initial value is ANGLE_MIN
            if LIMIT_TABLE_ANGLE (STATE, VALUE) then
               TABLE_STOP_H;
               STATE := SUCC (STATE)
            end if
         end loop
      end var
   end par
end process

--------------------------------------------------------

process P9 [G9: CRANE_POSITION,
            CRANE_TO_BELT2, CRANE_STOP_H, CRANE_TO_BELT1,
            DC_READY, DC, CF: NONE] is
   -- this process controls the position of the crane
   -- initial position is OTHER
   par CRANE_STOP_H in
      loop
         CRANE_TO_BELT2;
         CRANE_STOP_H;
         DC_READY;
         DC;
         CRANE_TO_BELT1;
         CRANE_STOP_H;
         CF
      end loop
    ||
      var STATE: TWO_STATE,
          VALUE: CRANE_POSITION
      in
         STATE := 1;
         loop
            G9 (?VALUE); -- initial value is OTHER
            if LIMIT_CRANE_POSITION (STATE, VALUE) then
               CRANE_STOP_H;
	       STATE := SUCC (STATE)
            end if
         end loop
      end var
   end par
end process

--------------------------------------------------------

process P10 [G10: CRANE_HEIGHT,
             CRANE_LIFT, CRANE_STOP_V, CRANE_LOWER, DC_READY, DC, CF: NONE] is
   -- this process controls the height of the crane
   -- initial height is OTHER
   par CRANE_STOP_V in
      loop
         DC_READY;
         CRANE_LOWER;
         CRANE_STOP_V;
         DC;
         CRANE_LIFT;
         CRANE_STOP_V;
         CF
      end loop
    ||
      var STATE: TWO_STATE,
          VALUE: CRANE_HEIGHT
      in
         STATE := 1;
         loop
            G10 (?VALUE); -- initial value is OTHER
            if LIMIT_CRANE_HEIGHT (STATE, VALUE) then
               CRANE_STOP_V;
               STATE := SUCC (STATE)
            end if
         end loop
      end var
   end par
end process

--------------------------------------------------------

process P11 [CRANE_MAG_ON, CRANE_MAG_OFF, DC, CF: NONE] is
   -- this process controls the magnet of the crane
   loop
      DC;
      CRANE_MAG_ON;
      CF;
      CRANE_MAG_OFF
   end loop
end process

--------------------------------------------------------

process P12 [G12: BOOL,
             BELT1_START, BELT1_STOP, BLANK_ADD, FT_READY, FT, CF: NONE] is
   -- this process controls belt 1 (feed belt)
   par
      BLANK_ADD ->
         BLANK_ADD;
         BLANK_ADD;
         BLANK_ADD;
         BLANK_ADD;
         BLANK_ADD;
         stop
    ||
      BLANK_ADD, BELT1_STOP, FT ->
         -- before the loop, a few actions are required because, initially,
         -- there is no blank in the production cell, so that a blank has
         -- first to be added and the feed belt has to be started to move
         -- this blank towards the elevating rotary table
         BLANK_ADD;
         BELT1_START;
         loop
            BELT1_STOP;
            FT_READY;
            BELT1_START;
            FT;
            select
               CF
             []
               BLANK_ADD
            end select
         end loop
    ||
      BELT1_STOP, FT ->
         var S13, PREVIOUS_S13: BOOL in
            PREVIOUS_S13 := false;
            loop
               G12 (?S13);
               if PREVIOUS_S13 != S13 then
                  if S13 then
                     BELT1_STOP
                  else
                     FT
                  end if;
                  PREVIOUS_S13 := S13
               end if
            end loop
         end var
   end par
end process

--------------------------------------------------------

process P13 [G13: BOOL,
             BELT2_START, BELT2_STOP, A2D, DC: NONE] is
   -- this process controls belt 2 (deposit belt)
   par BELT2_STOP in
      -- before the loop, a few actions are required because, initially,
      -- there is no item on the deposit belt, so that action DC would be
      -- impossible; the deposit belt has thus to wait for arm 2 to deliver
      -- an item (i.e., action A2D)
      A2D;
      loop
         BELT2_START;
         BELT2_STOP;
         par
            DC
          ||
            A2D
         end par
      end loop
    ||
      var S14, PREVIOUS_S14: BOOL in
         PREVIOUS_S14 := false;
         loop
            G13 (?S14);
            if PREVIOUS_S14 and not (S14) then
               BELT2_STOP
            end if;
            PREVIOUS_S14 := S14
         end loop
      end var
   end par
end process

end module
\end{lstlisting}

\subsection{Principal Module CELL}
\label{sec:lnt:cell}

This module defines the production cell controller as the parallel composition of the \T{CONTROLLER} and the (concurrent version of the) \T{DISPATCHER}.

\begin{lstlisting}
module CELL (TYPES, CHANNELS, CONTROLLER, DISPATCHER) is

process MAIN [GET_STATUS: STATUS,
              BLANK_ADD, 
              PRESS_UPWARD, PRESS_STOP, PRESS_DOWNWARD,
              ARM1_FORWARD, ARM1_STOP, ARM1_BACKWARD,
              ARM2_FORWARD, ARM2_STOP, ARM2_BACKWARD,
              ARM1_MAG_ON, ARM1_MAG_OFF,
              ARM2_MAG_ON, ARM2_MAG_OFF,
              ROBOT_LEFT, ROBOT_STOP, ROBOT_RIGHT,
              TABLE_LEFT, TABLE_STOP_H, TABLE_RIGHT,
              TABLE_UPWARD, TABLE_STOP_V, TABLE_DOWNWARD,
              CRANE_TO_BELT2, CRANE_STOP_H, CRANE_TO_BELT1,
              CRANE_LIFT, CRANE_STOP_V, CRANE_LOWER,
              CRANE_MAG_ON, CRANE_MAG_OFF,
              BELT1_START, BELT1_STOP,
              BELT2_START, BELT2_STOP: NONE] is
   hide
      G1: PRESS_POSITION,
      G2: ARM1_EXTENSION,
      G3: ARM2_EXTENSION, 
      G6: ROBOT_ANGLE,
      G7: TABLE_POSITION,
      G8: TABLE_ANGLE,
      G9: CRANE_POSITION,
      G10: CRANE_HEIGHT,
      G12, G13: BOOL
   in
      par G1, G2, G3, G6, G7, G8, G9, G10, G12, G13 in
         DISPATCHER [GET_STATUS, G1, G2, G3, G6, G7, G8, G9, G10, G12, G13]
                     (false) -- or true to sequentialize events
       ||
         CONTROLLER [G1, G2, G3, G6, G7, G8, G9, G10, G12, G13,
                     PRESS_UPWARD, PRESS_STOP, PRESS_DOWNWARD,
                     ARM1_FORWARD, ARM1_STOP, ARM1_BACKWARD,
                     ARM2_FORWARD, ARM2_STOP, ARM2_BACKWARD,
                     ARM1_MAG_ON, ARM1_MAG_OFF,
                     ARM2_MAG_ON, ARM2_MAG_OFF,
                     ROBOT_LEFT, ROBOT_STOP, ROBOT_RIGHT,
                     TABLE_LEFT, TABLE_STOP_H, TABLE_RIGHT,
                     TABLE_UPWARD, TABLE_STOP_V, TABLE_DOWNWARD,
                     CRANE_TO_BELT2, CRANE_STOP_H, CRANE_TO_BELT1,
                     CRANE_LIFT, CRANE_STOP_V, CRANE_LOWER,
                     CRANE_MAG_ON, CRANE_MAG_OFF,
                     BELT1_START, BELT1_STOP,
                     BELT2_START, BELT2_STOP,
                     BLANK_ADD]
      end par
   end hide
end process

end module
\end{lstlisting}

\end{document}